\documentstyle[12pt]{article}
\begin{document}
\renewcommand{\thetable}{\Roman{table}}
\newcommand{\be}{\begin{eqnarray}}
\newcommand{\beq}{\begin{equation}}
\newcommand{\ba}{\begin{array}}
\newcommand{\ee}{\end{eqnarray}}
\newcommand{\eeq}{\end{equation}}
\newcommand{\ea}{\end{array}}
\newcommand{\zt}{\zeta}
\newcommand{\ve}{\varepsilon}
\newcommand{\al}{\alpha}
\newcommand{\gm}{\gamma}
\newcommand{\Gm}{\Gamma}
\newcommand{\om}{\omega}
\newcommand{\et}{\eta}
\newcommand{\bt}{\beta}
\newcommand{\dt}{\delta}
\newcommand{\Dt}{\Delta}
\newcommand{\La}{\Lambda}
\newcommand{\la}{\lambda}
\newcommand{\vp}{\varphi}
\newcommand{\nn}{\nonumber}
\newcommand{\nid}{\noindent}
\newcommand{\lmx}[1]{\begin{displaymath} {#1}=
                    \left(\begin{array}{rrr}}
\newcommand{\rmx}{\end{array} \right) \end{displaymath}}

\begin{titlepage}
\begin{center}
 {\Large \bf
 On critical behavior of phase transitions in
 certain antiferromagnets with complicated ordering}

\vspace{1cm}
 {\Large A. I. Mudrov$^*$,  K. B. Varnashev$^{**}$}

\bigskip
{\it $^{*}$ Department of Mathematics, Bar-Ilan University,
52900 Ramat-Gan, Israel; \\
e-mail: mudrova@macs.biu.ac.il \\
$^{**}$
Department of Physical Electronics, Saint Petersburg
Electrotechnical University, \\
Professor Popov Street  5, St. Petersburg, 197376, Russia; \\
e-mail: kvarnash@kv8100.spb.edu}
\end{center}
\vspace{0.25cm}
\begin{abstract}
\vspace{0.75cm}
Within the four-loop $\ve$ expansion, we study the critical behavior of
certain antiferromagnets with complicated ordering. We show that an
anisotropic stable fixed point governs the phase transitions with new
critical exponents. This is supported by the estimate of critical
dimensionality $N_c^C=1.445(20)$ obtained from six loops via the exact
relation $N_c^C=\frac{1}{2} N_c^R$ established for the real and complex
hypercubic models.
\end{abstract}
\vspace{0.5cm}

\nid
~\qquad PACS{64.60.Ak, 64.60.Fr, 75.40.Cx}
\vspace{1cm}

\nid
~\qquad Published in: {\sl JETP Letters} {\bf 74}, No 5, 279-283 (2001)

\quad \qquad \qquad \qquad \leftline{[{\sl Pis'ma v Zh. Eksp. Teor. Fiz.} 
{\bf 74}, No 5, 309-313 (2001)]}

\vspace{8cm}

\rightline{{\sl Typeset using} \LaTeX}
\end{titlepage}

It is known that the critical properties of phase transitions in
certain antiferromagnets involving an increase of the unit cell
in one or more directions at the critical temperature can be described
by a generalized $2N$-component ($N \ge 2$)
Ginzburg-Landau model with three independent quartic terms
\be
\nn
H &=&
\int d^{~D}x \Bigl[\frac{1}{2} \sum_{i=1}^{2N} (m_0^2~ \vp_i^2
 + \vec \nabla \vp_i \vec \nabla \vp_i)
 + \frac{u_0}{4!} \Bigl(\sum_{i=1}^{2N} \vp_i^2 \Bigr)^2  \\
\nn \\
  &+& \frac{v_0}{4!} \sum_{i=1}^{2N} \vp_i^4
 + 2 \frac{z_0}{4!} \sum_{i=1}^N \vp_{2i-1}^2\vp_{2i}^2 \Bigr]
\label{eq:Ham}
\ee
associated with the isotropic, cubic, and tetragonal interactions,
respectively \cite{Muk}. Here $\vp_i$ is the real vector order parameter
in $D=4 - \ve$ dimensions and $m_0^2$ is proportional to the deviation
from the mean-field transition point.
When $N=2$, Hamiltonian (\ref{eq:Ham}) describes the antiferromagnetic
phase transitions in TbAu$_2$ and DyC$_2$ and the structural phase
transition in NbO$_2$ crystal\footnote{The phase transitions in helical magnets
Tb, Dy, and Ho belong to the same class of universality
\cite{Bak}.}. Another physically important case $N=3$
is relevant to the antiferromagnetic phase transitions in such substances as
K$_2$IrCl$_6$, TbD$_2$, MnS$_2$, and Nd. All these phase transitions are
known from experiments to be of second order\footnote{An interesting
type of multisublattice antiferromagnets, such as MnO, CoO, FeO, and NiO, was
studied in Ref. \cite{Dz,Bak1,Muk1}. It was shown, in the leading orders in
$\ve$, that the phase transitions in these substances are of first order.}
(see Ref. \cite{TMTB} and references
therein). However, the experimental data were insufficiently accurate
to provide reliable values of critical exponents and the obtained estimates
\cite{Als,Pynn,Sch} were found to differ significantly from the theoretically
expected numbers.

For the first time the magnetic and structural phase transitions described
by model (\ref{eq:Ham}) were studied in the framework of the renormalization
group (RG) by Mukamel and Krinsky within the lowest orders in $\ve$
\cite{Muk,Muk1}. A three-dimensionally stable fixed point (FP) with
coordinates $u^* > 0$, $v^* = z^* > 0$ was predicted\footnote{Following Mukamel
\cite{Muk}, we call this point "unique".}. That point was shown to
determine a new universality class with a specific set of critical exponents.
However, for the physically important case $N=2$, the critical exponents of
this unique stable FP turned out to be exactly the same as those of the $O(4)$\d
symmetric one.

For the years an alternative analysis of critical behavior of the model, the
RG approach in three dimensions, was carried out within the two- and
three-loop approximations \cite{Shp,VS}. Those investigations gave the same
qualitative predictions: the unique stable FP does exist on the 3D
RG flow diagram. By using different resummation procedures, the critical
exponents computed at this point proved to be close to those of the Bose
FP ($u=0$, $v=z>0$) rather than the isotropic
($O(N)$-symmetric; $u>0$, $v=z=0$) one.
It was also shown that the unique and the
Bose FPs are very close to each other, so that they may interchange
their stability in the next orders of RG approximation \cite{VS}.

Recently, the critical properties of the model were
analyzed in third order in $\ve$ \cite{BGMV,MV-1}. Investigation of
the FP stability and calculation of the critical dimensionality
$N_c$ of the order parameter, separating two different regimes of critical
behavior\footnote{When $N>N_c$ the unique FP is stable in 3D
while for $N<N_c$ the stable FP is the isotropic one.},
confirmed that model (\ref{eq:Ham}) has the unique stable
FP at $N=2$ and $N=3$.
However, the twofold degeneracy of the stability matrix eigenvalues at
the one-loop level was observed for this FP \cite{MV-1}.
That degeneracy was shown to cause a substantial decrease of the accuracy
expected within the three-loop approximation and powers of $\sqrt{\ve}$ to
appear in the expansions\footnote{Similar phenomenon was observed earlier
in studying the impure Ising model
(see Refs. \cite{Khm}). Half-integer powers in $\ve$ arising in that model
have different origin but also lead to the loss of accuracy.}.
So, computational difficulties were shown to grow faster than the
amount of essential information one may extract from high-loop
approximations. That resulted in the conclusion that the $\ve$-expansion
method is not quite effective for the given model.

Another problem associated with model (\ref{eq:Ham}) is the question whether
the unique FP is really stable in 3D, thus leading to a new class of
universality, or its stability is only an effect of insufficient accuracy of
the RG approximations used. Indeed, there are general nonperturbative
theoretical arguments indicating that the only stable FP in 3D may be
the Bose one and the phase transitions of interest should be governed by that
stable FP \cite{CB}. However, up to now this assertion found no
confirmation within the RG approach. In such a situation it is highly
desirable to extend already known $\ve$ expansions for the stability matrix
eigenvalues, critical exponents, and the critical dimensionality in order to
apply more sophisticated resummation technique to longer expansions.

In this Letter we, firstly, avoid the problem of the eigenvalues
degeneracy in model (\ref{eq:Ham}) by analyzing the critical behavior
of an equivalent complex $N^C$-component order parameter model
with the effective Hamiltonian
\be
H =
\int d^D x \Bigl[\frac{1}{2}( m_0^2 \psi_i\psi^*_i
 + \vec {\nabla} \psi_i \vec {\nabla} \psi^*_i)
 + \frac{u_0}{4!}\ \psi_i \psi^*_i \psi_j \psi^*_j
 + \frac{v_0}{4!}\ \psi_i \psi_i \psi^*_i \psi^*_i \Bigr]
\label{eq:Ham1}
\ee
comprising the isotropic and cubic interactions\footnote{The model
with the complex vector order parameter was considered by 
Dzyaloshinskii \cite{Dz77} in studying the phase transitions in
DyC$_2$, TbAu$_2$ ($N^C=2$) and TbD$_2$, MnS$_2$, and Nd
($N^C=3$).}. Note that this Hamiltonian
also describes the  real  hypercubic  model  \cite{Ah}  if  $\psi_i$
is thought to be the real $N^R$-component order parameter.
The model (\ref{eq:Ham1}) comes out exactly from model (\ref{eq:Ham}) at
$v_0=z_0$ and it is free from the eigenvalues degeneracy.
Secondly, we examine the existence of the anisotropic stable FP
in model (\ref{eq:Ham1}) on the basis of the higher-order
$\ve$ expansion. Namely, using dimensional regularization and the minimal
subtraction scheme \cite{Hooft}, we derive the four-loop RG functions as
power series in $\ve$ and analyze the FP stability.
For the first time, we give realistic numerical estimates for the stability
matrix eigenvalues using the Borel transformation with a conformal mapping
\cite{LgZ}. This allows us to carry out the careful analysis of the stability
of all the FPs of the model. We state the exact relation
$N_c^C=\frac{1}{2} N_c^R$ between the critical (marginal) spin
dimensionalities of the real and complex hypercubic models and obtain the
estimate $N_c^C=1.445(20)$ using six-loop results of Ref. \cite{CPV}.
We show that the anisotropic (complex cubic; $u\ne0$, $v\ne0$) stable FP
of model (\ref{eq:Ham1}), being the counterpart of the unique point in
model (\ref{eq:Ham}), does
exist on 3D RG flow diagram at $N^C > N_c^C$. For this stable FP we
give more accurate critical exponents estimates in comparison with the
previous three-loop results \cite{MV-1} by applying the summation technique
of Ref. \cite{MV-2} to the longer series.

The four-loop $\ve$ expansion for the $\bt$-functions of model
(\ref{eq:Ham1}) were recently obtained by the present authors in Ref.
\cite{MV00p}. From the system of equations $\bt_u(u^*, v^*)=0$,
$\bt_v(u^*, v^*)=0$ one can calculate formal series for the four FPs:
the trivial Gaussian one and nontrivial isotropic, Bose, and complex cubic
FPs. Instead of  presenting here the FPs themselves, which
have no direct physical meaning, we present the eigenvalues of the stability
matrix
\begin{equation}
\Omega  \> = \>
\left(
\begin{array}{cc}
    \frac{\partial\bt_u(u,v)}{\partial u} &
    \frac{\partial\bt_u(u,v)}{\partial v} \\
    \frac{\partial\bt_v(u,v)}{\partial u} &
    \frac{\partial\bt_v(u,v)}{\partial v}
\end{array}
\right)
\label{eq:Om}
\end{equation}
taken at the most intriguing Bose and complex cubic FPs.
They are
\be
\nn
\om_1 &=& - \frac{1}{2} \ve + \frac {6}{20} \ve^2
+ \frac{1}{8} \Biggl[- \frac {257}{125}
- \frac {384}{125} \zt(3) \Biggr] \ve^3 \\
\nn
      &+& \frac{1}{16} \Biggl[\frac {5109}{1250}
+ \frac {624}{125} \zt(3)
- \frac {576}{125} \zt(4)
+ \frac {3648}{125} \zt(5) \Biggr] \ve^4 ,
\ee
\be
\nn
\om_2 &=& \frac {1}{10} \ve - \frac {14}{100} \ve^2
+ \frac{1}{8} \Biggl[- \frac {311}{625}
+ \frac {768}{625} \zt(3) \Biggr] \ve^3 \\
      &+& \frac{1}{16} \Biggl[- \frac {61}{250}
+ \frac {3752}{3125} \zt(3)
+ \frac {1152}{625} \zt(4)
- \frac {4864}{625} \zt(5) \Biggr] \ve^4
\label{eq:Bose}
\ee
at the Bose FP and for $N^C=2$
\be
\nn
\om_1 &=& - \frac{1}{2} \ve
  + \frac{13}{48} \ve^2 + \frac{1}{8} \Biggl[- \frac {65}{36}
- \frac {7}{3} \zt(3) \Biggr] \ve^3 \\
\nn
      &+& \frac{1}{16} \Biggl[\frac {1679}{432}
  + \frac {169}{36} \zt(3)
  - \frac {7}{2} \zt(4)
  + \frac {365}{18} \zt(5) \Biggr] \ve^4 ,
\ee
\be
\nn
\om_2 &=& - \frac{1}{12} \ve^2
  + \frac{1}{8} \Biggl[ \frac {5}{18}
  + \frac {5}{6} \zt(3) \Biggr] \ve^3 \\
      &+& \frac{1}{16} \Biggl[\frac {181}{144}
  - \frac {145}{72} \zt(3)
  + \frac {5}{4} \zt(4)
  - \frac {50}{9} \zt(5) \Biggr] \ve^4
\label{eq:CC2}
\ee
and $N^C=3$
\be
\nn
\om_1 &=& - \frac{1}{2} \ve
  + \frac{58}{220} \ve^2 + \frac{1}{8} \Biggl[- \frac {19533}{15125}
- \frac {14832}{6655} \zt(3) \Biggr] \ve^3 \\
\nn
      &+& \frac{1}{16} \Biggl[\frac {310518757}{91506250}
  + \frac {1644864}{1830125} \zt(3)
  - \frac {22248}{6655} \zt(4)
  + \frac {283056}{14641} \zt(5) \Biggr] \ve^4 ,
\ee
\be
\nn
\om_2 &=& - \frac{1}{22} \ve
  + \frac{2}{2420} \ve^2 + \frac{1}{8} \Biggl[\frac {90363}{166375}
- \frac {3408}{73205} \zt(3) \Biggr] \ve^3 \\
      &+& \frac{1}{16} \Biggl[\frac {1151231173}{1006568750}
  - \frac {50696504}{20131375} \zt(3)
  - \frac {5112}{73205} \zt(4)
  + \frac {107344}{161051} \zt(5) \Biggr] \ve^4
\label{eq:CC3}
\ee
at the complex cubic one, where $\zt(3)$, $\zt(4)$, and $\zt(5)$ are
the Riemann $\zt$ functions.

It is known that RG series are at best asymptotic.
An appropriate resummation procedure has to be applied in order to extract
reliable physical information from them. To obtain the eigenvalue
estimates we have used an approach based on the Borel transformation
modified with a conformal mapping \cite{LgZ,MV-2}. If both eigenvalues of
matrix (\ref{eq:Om}) are
negative, the associated FP is infrared stable and the critical
behavior of experimental systems undergoing second-order phase transitions
is determined only by that stable point. For the Bose and
the complex cubic FPs our numerical results are presented in Table I.
It is seen that the complex cubic FP is absolutely stable in $D=3$
($\ve=1$), while the Bose point appears to be of the "saddle" type.
However $\om_2$'s of either points are very small at the four-loop level,
thus implying that these points may swap their stability in the next order
of RG approximation. We can compare $\om_2$  at the
complex cubic FP quoted in Table I with the three-loop results of
Ref. \cite{Shp} obtained in the framework of RG approach directly
in 3D. Those estimates $\om_2=-0.010$ for $N^C=2$ and $\om_2=-0.011$
for $N^C=3$
are solidly consistent with ours.

The four-loop $\ve$ expansion for the critical dimensionality of the order
parameter of model (\ref{eq:Ham1}) reads
$$N_c^C = 2 - \ve + \frac{5}{24} \Bigl[6 \zt(3) -1 \Bigr] \ve^2
 + \frac{1}{144} \Bigl[45 \zt(3)
 + 135 \zt(4) - 600 \zt(5) -1 \Bigr] \ve^3 .$$
Instead of processing this expression numerically, we state the
exact relation $N_c^C=\frac{1}{2} N_c^R$, which is independent on the order
of approximation used.
In fact, the critical dimensionality $N_c^C$ for the complex cubic model is
determined as that value of $N^C$, at  which the complex cubic FP
coincides with the isotropic one. In the same way, the
critical dimensionality $N_c^R$ is defined for the real cubic model.
Both systems exhibit effectively the isotropic critical behavior at
$N^C=N^C_c$ and $N^R=N^R_c$. So, because the complex $O(2N^C)$-symmetric
model is equivalent to the real $O(N^R)$-symmetric one, the relation
$2 N_c^C=N_c^R$ holds true. For $N^C>N_c^C$ the complex cubic FP
of model (\ref{eq:Ham1}) should be stable in 3D.

The five-loop $\ve$ expansion for $N_c^R$ was recently obtained in Ref.
\cite{KSf}.
Resummation of that series gave the estimate $N_c^R=2.894(40)$ (see Ref.
\cite{VKB}).
Therefore we conclude that $N_c^C=1.447(20)$ from the five-loops.
Practically the same estimate $N_c^C=1.435(25)$ follows from a
constrained analysis of $N_c^R$ taking into account $N_c^R=2$ in two
dimensions \cite{CPV}.
From the recent pseudo-$\ve$ expansion
analysis of the real hypercubic model \cite{FHY} one can extract
$N_c^C=1.431(3)$. However the most accurate estimate $N_c^C=1.445(20)$ results
from  the value $N_c^R=2.89(4)$ obtained on the basis of the numerical
analysis of the four-loop \cite{VKB} and the six-loop
\cite{CPV} 3D RG expansions for the $\bt$-functions of the real hypercubic
model.

Finally, we have computed the four-loop $\ve$ series for the critical
exponents. At the stable complex cubic FP they are
\be
\nn
\eta &=&
     \frac{\ve^2}{48} + \frac{5}{288} \ve^3
      - \frac{21 \zt(3) - 13}{1728} \ve^4 ,
\ee
\be
\nn
\gm^{-1}
&=& 1 - \frac{\ve}{4} - \frac{7}{96} \ve^2
      + \frac{84 \zt(3) - 1}{1152} \ve^3 \\
      &-& \frac{1420 \zt(3) - 1512 \zt(4)
      + 5840 \zt(5) - 2059}{27648} \ve^4
\label{eq:EGr-2}
\ee
for $N^C=2$ and
\be
\nn
\eta &=&
     \frac{5}{242} \ve^2 + \frac{177}{10648} \ve^3
      - \frac{59328 \zt(3) - 50083}{5153632} \ve^4 ,
\ee
\be
\nn
\gm^{-1}
&=& 1 - \frac{3}{11} \ve - \frac{7}{242} \ve^2
      + \frac{912 \zt(3) + 3905}{58564} \ve^3 \\
      &-& \frac{207682 \zt(3) - 15048 \zt(4)
      + 30320 \zt(5) - 151817}{1288408} \ve^4
\label{eq:EGr-3}
\ee
for $N^C=3$.
Other critical exponents can be found through the known scaling relations.
The numerical estimates obtained are collected in Table II.
The critical exponents for the isotropic
and the Bose FPs are also presented, for comparison. We can compare our
results with the available experimental data. For example, in the case of
the structural transition in the NbO$_2$ crystal the critical exponent of
spontaneous polarization was measured in Ref. \cite{Pynn}, $0.33< \bt < 0.44$.
Our estimate $\bt=0.371$ obtained using the data of Table II and scaling
relations lies in that interval.

In summary, the four-loop $\ve$-expansion analysis of the Ginzburg\d
Landau model with cubic anisotropy and complex vector order parameter
relevant to the phase transitions in certain antiferromagnets with
complicated ordering has been carried out. Investigation of the global
structure of RG flows for the physically significant cases $N^C=2$ and
$N^C=3$
leads to the conclusion that the anisotropic complex
cubic FP is absolutely stable in 3D. Therefore the critical
thermodynamics of the phase transitions in the NbO$_2$ crystal and
in the antiferromagnets TbAu$_2$, DyC$_2$, K$_2$ IrCl$_6$, TbD$_2$, MnS$_2$, 
and Nd should govern by this stable point with a specific set of critical
exponents, in the frame of the given approximation. The critical dimensionality
$N_c^C=1.445(20)$ obtained from six loops supports this conclusion.
At the complex cubic FP, the critical exponents were calculated
using the Borel summation technique in combination with a conformal
mapping. For the structural phase transition in NbO$_2$ and for the
antiferromagnetic phase transitions in TbAu$_2$ and DyC$_2$ they were shown
to be close to the critical exponents of the $O(4)$-symmetric model.
In contrast to this, the critical exponents for the antiferromagnetic phase
transitions in K$_2$IrCl$_6$, TbD$_2$, MnS$_2$, and Nd turned out to be 
close to the Bose ones.

Although our calculations show that the complex cubic FP,
rather than the Bose one, is stable at the four-loop level, the
eigenvalues $\om_2$ of both points are very small.
Therefore the situation is very close to marginal, and the FPs might
change their stability to opposite in the next order of
perturbation theory, so that the Bose point would occur stable.
This conjecture is in agreement with the recent six-loop RG study of
three-coupling-constant model (\ref{eq:Ham}) directly in three dimensions
\cite{PV-z}.
The authors argue the global stability of the Bose FP, although
the numerical estimate  $\om_2=-0.007(8)$ of the smallest stability matrix
eigenvalue of the Bose point appears to be very small and
the apparent accuracy of the analysis does not exclude the opposite sign for
$\om_2$. In this situation it would be very desirable to compare
the critical exponents values obtained theoretically with values that could
be determined from experiments, in order to verdict which of the two FPs
is really stable in physical space.
Finally, it would be also useful to investigate certain universal amplitude
ratios of the model because they vary much more among different universality
classes than exponents do and might be more effective as a diagnostic tool.

We are grateful to Professor M. Henkel for helpful remarks and to
Dr. E. Vicari for sending to us a copy of the review cited in Ref. \cite{PV-z}.
This work was supported by the Russian Foundation for Basic Research via grant
No. 01-02-17048, and by the Ministry of Education of Russian Federation via
grant No. E00-3.2-132.

\begin{table}
\caption{Eigenvalue exponent estimates for the Bose (BFP) and the complex
cubic (CCFP) FPs at $N^C=2$ and $N^C=3$ obtained in the four-loop
order in $\ve$ ($\ve=1$) using the Borel transformation with a conformal
mapping.}
\label{TabI}
\vspace{0.5cm}
\hspace{1.5cm}
\begin{tabular}{|c|l|l|l|l|}\hline
Type of             &\multicolumn{2}{|c|}{$N^C=2$}&
                     \multicolumn{2}{|c|}{$N^C=3$}
                   \\[0pt]    \cline{2-5}
FP         &\multicolumn{1}{|c|}{$\om_1$}
                    &\multicolumn{1}{|c|}{$\om_2$}
                    &\multicolumn{1}{|c|}{$\om_1$}
                    &\multicolumn{1}{|c|}{$\om_2$}
                   \\[0pt] \hline
BFP                 &$-0.395(25)$ & $0.004(5)$
                    &$-0.395(25)$ & $0.004(5)$
                   \\[0pt] \hline
CCFP                &$-0.392(30)$ & $-0.029(20)$
                    &$-0.400(30)$ & $-0.015(6)$
                   \\[0pt]\hline
\end{tabular}
\end{table}

\begin{table}
\caption{Critical exponents for the isotropic (IFP), the Bose
(BFP), and the complex cubic (CCFP) FPs at $N^C=2$ and $N^C=3$
calculated in the four-loop order in $\ve$ ($\ve=1$) using the Borel
transformation with a conformal mapping.}
\label{TabII}
\vspace{0.5cm}
\begin{tabular}{|c|l|l|l|l|l|l|}\hline
Type of             &\multicolumn{3}{|c|}{$N^C=2$}&
                     \multicolumn{3}{|c|}{$N^C=3$}
                   \\[0pt]    \cline{2-7}
FP         &\multicolumn{1}{|c|}{$\eta$}
                    &\multicolumn{1}{|c|}{$\nu$}
                    &\multicolumn{1}{|c|}{$\gm$}
                    &\multicolumn{1}{|c|}{$\eta$}
                    &\multicolumn{1}{|c|}{$\nu$}
                    &\multicolumn{1}{|c|}{$\gm$}
                   \\[0pt]  \hline
IFP                 &$0.0343(20)$ & $0.725(15)$ & $1.429(20)$
                    &$0.0317(10)$ & $0.775(15)$ & $1.524(25)$
                   \\[0pt] \hline
BFP                 &$0.0348(10)$ & $0.664(7)$ & $1.309(10)$
                    &$0.0348(10)$ & $0.664(7)$ & $1.309(10)$
                   \\[0pt] \hline
CCFP                &$0.0343(20)$ & $0.715(10)$ & $1.404(25)$
                    &$0.0345(15)$ & $0.702(10)$ & $1.390(25)$
                   \\[0pt]\hline
\end{tabular}
\end{table}

\newpage

\nid

\end{document}